\documentclass[aps,twocolumn,floatfix,amsmath,amsfonts]{revtex4-1}

\usepackage{graphicx}
\usepackage{braket}
\usepackage{bm}
\usepackage{xcolor}
\usepackage{upgreek}
\usepackage{tabularx}
\usepackage[normalem]{ulem}
\usepackage[utf8]{inputenc}

\begin{document}
\title{Green exciton series in cuprous oxide}
\author{Patric Rommel}
\email[Email: ]{patric.rommel@itp1.uni-stuttgart.de}
\author{Patrik Zielinski}
\author{J\"org Main}
\affiliation{Institut f\"ur Theoretische Physik 1, Universit\"at
  Stuttgart, 70550 Stuttgart, Germany}

\date{\today}

\begin{abstract}
We numerically investigate the odd parity states of the green exciton
series in cuprous oxide. Taking into account the coupling to the
yellow series and especially to the yellow continuum, the green
excitons are quasi-bound resonances with a finite lifetime which
cannot be described with Hermitian operators.  To calculate their
positions and linewidths, we use the method of complex-coordinate
rotation, leading to a non-Hermitian complex eigenvalue problem.
We find that the behavior of the dominant P states is very well
approximated by a modified Rydberg formula using a negative quantum
defect. The corresponding linewidths induced by the coupling to the
yellow continuum decrease with the third power of the principal
quantum number.
\end{abstract}

\maketitle

\section{Introduction}
\label{sec:introduction}
Early experiments on cuprous oxide in the 1950s were able to observe
excitons with principal quantum numbers up to $n = 9$
\cite{Gross1956}.
Since then, experimental methods have made tremendous progress.
After Kazimierczuk \emph{et al.}\ realized principal quantum numbers
up to $n=25$ for the yellow series in cuprous
oxide~\cite{GiantRydbergExcitons}, interest has been drawn to the
field of giant Rydberg excitons.
Due to the influence of the crystal symmetry, the exciton sequence
shows deviations from a perfect hydrogen-like spectrum.
For example, a splitting of the P and F states is observable
\cite{ObservationHighAngularMomentumExcitons,ImpactValence}.

In an idealized model, excitons can be described as bound states
between electrons and holes.
In general, a range of mechanisms make scattering possible and induce
a finite linewidth.
The excitons become quasi-bound states or resonances.
In case of the yellow excitons, the most prominent process is the
scattering with phonons \cite{franklinewidth,Stolz2018}.
Bound states can also become resonances by application of an external
electric field, which allows for tunneling processes into the unbound
region \cite{Heck18,Zielinski2019}.

In cuprous oxide, the excitons constituted by electrons in the lowest
conduction $\Gamma_6^+$ band and holes in the highest $\Gamma_7^+$
valence band are part of the yellow exciton series.
An electron can also be lifted from the $\Gamma_8^+$ valence band into
the $\Gamma_6^+$ conduction band, forming a green
exciton~\cite{Nikitine1959,Grun1961,Nikitine1963}.
Since the energy of the $\Gamma_8^+$ valence band is lowered by an
amount $\Delta$ in comparison to the uppermost $\Gamma_7^+$ valence
band, all green excitons, except for the even parity 1S states
investigated in \cite{Uihlein1981,frankevenexcitonseries},
lie within the energy range of the yellow continuum.
Yellow and green states are coupled by the valence band
structure~\cite{Luttinger52CyclotronResonanceSemiconductors,ImpactValence},
and the green states {with principal quantum numbers $n\ge 2$}
are therefore resonances instead of truly bound states.

{An efficient numerical method for the computation of the bound
states of Cu$_2$O including the impact of the valence band structure
(but ignoring the phonon coupling) is the diagonalization of the
Hamiltonian using a complete basis set \cite{ImpactValence,frankevenexcitonseries}.
The method can be applied to obtain the bound states of the yellow
exciton series at energies below the gap energy $E_{\mathrm{g}}=2.17208\,$eV and the
green 1S excitons, which are the only bound states of the green
exciton series.
However, the Hermitian eigenvalue problem does not allow for the
computation of unbound resonance states.}

Recent work by Kr\"uger and Scheel has focused on the interseries
transitions such as between the yellow and green excitons
\cite{KruegerInterseries2019}.
A better understanding of the {unbound resonances of the} green
series is thus of interest.
A convenient description of these resonance states is
achieved by the introduction of a complex energy, where the imaginary
part is related to the linewidth of the quasi-bound state.
These complex energies can be calculated by way of the
complex-coordinate{-}rotation method \cite{Reinhardt1982,Ho83,MOISEYEV1998},
where a complex scaling operation is performed to expose the resonance
positions in the complex plane.
{In this paper the numerical algorithm introduced in
Refs.~\cite{ImpactValence,frankevenexcitonseries} for the computation
of bound excitons is augmented by application of the complex-coordinate rotation.
This rotation turns the Hermitian eigenvalue problem
into a non-Hermitian system, and thus allows, for the first time, for
the computation of the complex resonance energies of the green
excitons.}
To this end, we first introduce the necessary theory in
Sec.~\ref{sec:theory}, including the method of complex-coordinate 
rotation and the exciton Hamiltonian, also giving a short
discussion of the numerical diagonalization and the extraction of the
oscillator strengths.
In Sec.~\ref{sec:results} we present our numerical results and discuss
their implications.
Finally, we draw conclusions and give a brief outlook in
Sec.~\ref{sec:conclusion}.

\section{Theory}
\label{sec:theory}
%
Both the yellow and the green series in Cu$_2$O have a unified
description in terms of the
Hamiltonian~\cite{ImpactValence,Schoene2016,Luttinger52CyclotronResonanceSemiconductors}
\begin{equation}
  H = E_{\rm g}+H_{\rm e}(\boldsymbol{p}_{\rm e})+H_{\rm h}(\boldsymbol{p}_{\rm h})
  +V(\boldsymbol{r}_{\rm e}-\boldsymbol{r}_{\rm h}) \, .
\label{eq:hamiltonian}
\end{equation}
Here, $E_\mathrm{g}$ denotes the gap energy, $H_{\rm e}$ and $H_{\rm h}$
are the electron and hole kinetic energies, and $V$ is the screened
Coulomb potential
\begin{equation}
  V(\boldsymbol{r}_{\rm e}-\boldsymbol{r}_{\rm h}) =
  -\frac{e^2}{4\pi\varepsilon_0\varepsilon|\boldsymbol{r}_{\rm e}-\boldsymbol{r}_{\rm h}|}\,,
\end{equation}
with the dielectric constant $\varepsilon$.
The electron and hole kinetic energies are given by
\begin{align}
  H_{\rm e}(\boldsymbol{p}_{\rm e}) &= \frac{\boldsymbol{p}_{\rm e}^2}{2m_{\rm e}} \, , \\
  H_{\rm h}(\boldsymbol{p}_{\rm h}) &= H_{\rm SO}+\frac{1}{2\hbar^2m_0}
  \{\hbar^2(\gamma_1+4\gamma_2)\boldsymbol{p}^2_{\rm h}\nonumber\\
&+2(\eta_1+2\eta_2)\boldsymbol{p}^2_{\rm h}(\boldsymbol{I}\cdot\boldsymbol{S}_{\rm h})\nonumber\phantom{\frac{1}{2}}\\
&-6\gamma_2(p^2_{\rm h1}\boldsymbol{I}^2_1+{\rm c.p.})
-12\eta_2(p^2_{\rm h1}\boldsymbol{I}_1\boldsymbol{S}_{\rm h1}+{\rm c.p.})\nonumber\phantom{\frac{1}{2}}\\
&-12\gamma_3(\{p_{\rm h1},p_{\rm h2}\}\{\boldsymbol{I}_1,\boldsymbol{I}_2\}+{\rm c.p.})\nonumber\phantom{\frac{1}{2}}\\
\phantom{\frac{1}{2}} &-12\eta_3(\{p_{\rm h1},p_{\rm h2}\}(\boldsymbol{I}_1\boldsymbol{S}_{\rm h2}
         +\boldsymbol{I}_2\boldsymbol{S}_{\rm h1})+{\rm c.p.})\} \, ,
\end{align}
with $m_{\rm e}$ and $m_0$ the effective and free electron mass,
respectively, $\{a,b\}=\frac{1}{2}(ab+ba)$ the symmetrized product,
and c.p.\ denotes cyclic permutation.
The spin-orbit interaction
\begin{equation}
  H_{\rm SO}=\frac{2}{3}\Delta
  \left(1+\frac{1}{\hbar^2}\boldsymbol{I}\cdot\boldsymbol{S}_{\rm h}\right)\,,
\end{equation}
couples the hole spin $\boldsymbol{S}_{\rm h}$ and the quasispin
$\boldsymbol{I}$ introduced to describe the degeneracy of the valence
band Bloch functions.
The yellow and green series are split by the energy $\Delta$.
The parameters $\eta_j$ and the three Luttinger parameters $\gamma_j$
parameterize the influence of the complex band structure and the
deviation from a parabolic dispersion relation.
In contrast to Refs.~\cite{frankevenexcitonseries,frankjanpolariton}
we here focus on odd excitons and thus can neglect the central-cell
corrections.
Finally, the Hamiltonian \eqref{eq:hamiltonian} is transformed into
relative and center-of-mass coordinates~\cite{Schmelcher1992},
\begin{eqnarray}
  \boldsymbol{r}&= \boldsymbol{r}_{\rm e}-\boldsymbol{r}_{\rm h}\, ,\quad
  \boldsymbol{R}=\frac{m_{\rm h}\boldsymbol{r}_{\rm h}+m_{\rm e}\boldsymbol{r}_{\rm e}}{m_{\rm h}+m_{\rm e}}\, ,\nonumber\\
  \boldsymbol{P}&= \boldsymbol{p}_{\rm e}+\boldsymbol{p}_{\rm h}\, ,\quad
  \boldsymbol{p}=\frac{m_{\rm h}\boldsymbol{p}_{\rm e}-m_{\rm e}\boldsymbol{p}_{\rm h}}{m_{\rm h}+m_{\rm e}} \, ,
\end{eqnarray}
with vanishing center-of-mass momentum $\boldsymbol{P} = 0$.
The material parameters for Cu$_2$O are given in Table~\ref{tab:MaterialParamters}.
\begin{table}[b]
\renewcommand{\arraystretch}{1.2}
     \centering
     \caption{Material parameters of Cu$_2$O used in the calculations.}
     \begin{tabular}{l|l c}
     \hline
       Energy gap  & $E_{\rm g}=2.17208\,$eV & \cite{GiantRydbergExcitons}\\
       Spin-orbit coupling       & $\Delta=0.131\,$eV &\cite{SchoeneLuttinger}\\
       Effective electron mass  & $m_{\rm e}=0.99m_0$ & \cite{HodbyEffectiveMasses} \\
       Effective hole mass  & $m_{\rm h}=0.58m_0$ & \cite{HodbyEffectiveMasses} \\
       Dielectric constant      & $\varepsilon=7.5$ &\cite{LandoltBornstein1998DielectricConstant}\\
       Valence band parameters & $\gamma_1=1.76$&\cite{SchoeneLuttinger}\\
       ~~& $\gamma_2=0.7532$&\cite{SchoeneLuttinger}\\
       ~~& $\gamma_3=-0.3668$&\cite{SchoeneLuttinger}\\
        ~~& $\eta_1=-0.020$&\cite{SchoeneLuttinger}\\
        ~~& $\eta_2=-0.0037$&\cite{SchoeneLuttinger}\\
        ~~& $\eta_3=-0.0337$&\cite{SchoeneLuttinger}\\
     \hline
     \end{tabular}
\label{tab:MaterialParamters}
\end{table}

\begin{figure*}
  \includegraphics[width=1.0\textwidth]{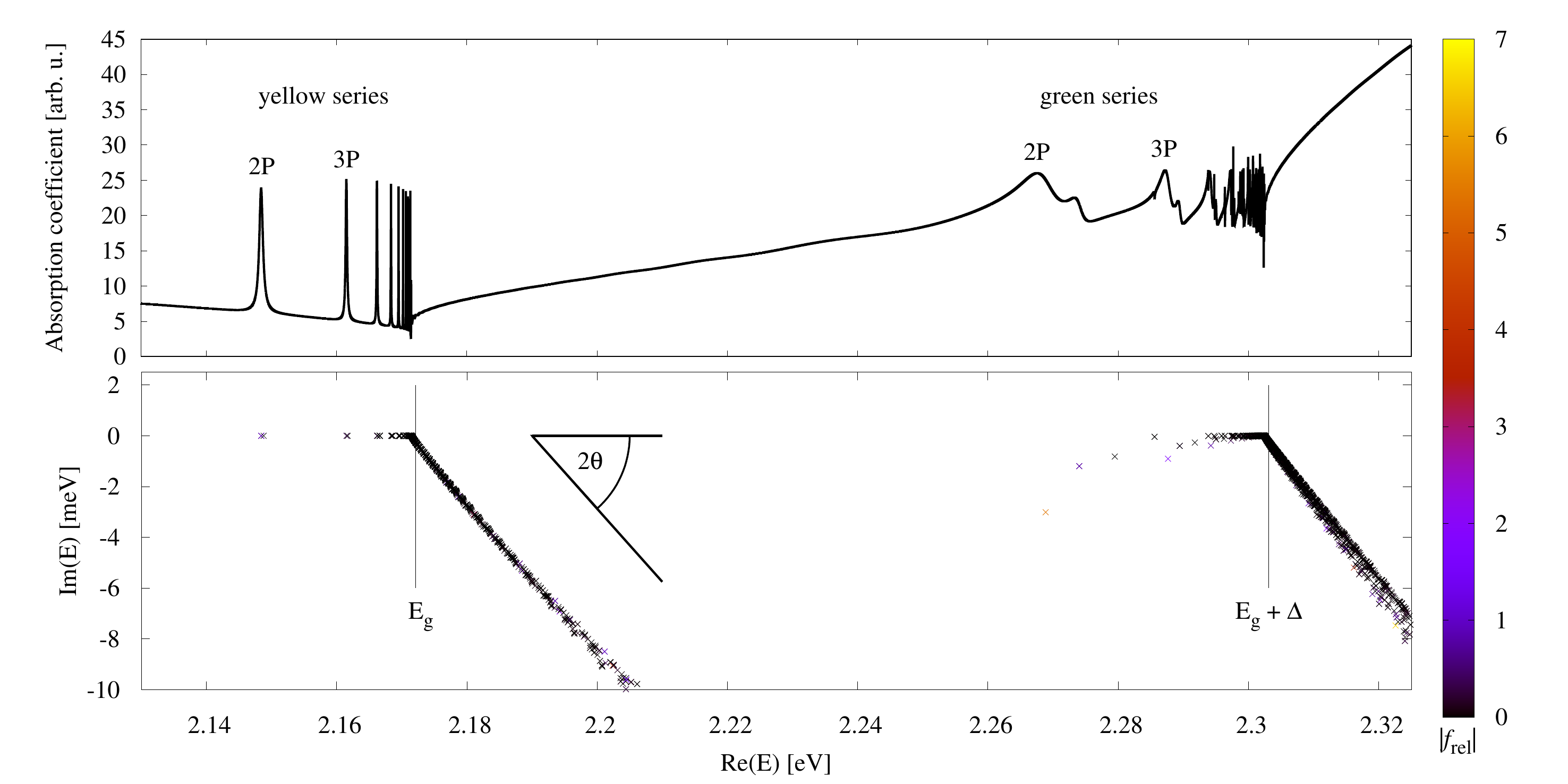}
  \caption{Spectrum and resonance positions in the complex plane for
    the yellow and green exciton series of Cu$_2$O.  Note that only
    the odd parity states are included.  The color bar shows the
    absolute value of the relative oscillator strength $|f_\mathrm{rel}|$.
    As coupling with phonons is not considered here, the linewidths of
    the yellow exciton states in the upper panel were put in by hand,
    as explained in the text. The vertical lines mark the respective
    band gaps.  Due to the finite basis, the numerical resonance
    positions are already rotated into the lower complex plane at a
    slightly lower energies.  The absorption coefficient is given in
    arbitrary units and with an arbitrary shift of the base line.
    \label{fig:SpectrumResonances}}
\end{figure*}
The yellow exciton series in cuprous oxide can now be investigated as
in our previous work~\cite{ImpactValence,frankevenexcitonseries} by 
diagonalizing the resulting Hamiltonian~\eqref{eq:hamiltonian}.
States lying above the band gap energy $E_{\mathrm{g}}=2.17208\,$eV on
the other hand are resonances not accessible by this method.
For their study, in this paper we augment the approach by the
complex-coordinate rotation~\cite{Reinhardt1982,Ho83,MOISEYEV1998},
allowing for the calculation of the complex energies of the green
exciton states.
The wave functions $\psi$ associated with resonances are not
normalizable and thus are not part of the Hilbert space.
They fulfill the Schr\"odinger equation
\begin{equation}
  H\psi = E \psi
  \label{UnrotatedSchroedinger}
 \end{equation}
with a complex energy $E$. 
We introduce the complex scaling operation $S_\theta$ with
\begin{equation}
 S_\theta \psi (\boldsymbol{r}) = \psi (\boldsymbol{r}\mathrm{e}^{\mathrm{i}\theta})\,.
 \label{eq:complexScaling}
\end{equation}
{Note that formally the operator $S_\theta$ can be applied to a
state $|\psi\rangle$ in arbitrary representation, however, Eq.~\eqref{eq:complexScaling}
is only valid in coordinate representation as used throughout this
paper.}
For sufficiently large $\theta$, $S_\theta\psi$ becomes square integrable~\cite{MOISEYEV1998}.
The rotated Schr\"odinger equation is given by
\begin{equation}
 S_\theta HS_\theta^{-1} S_\theta \psi = E S_\theta\psi\,.
 \label{RotatedSchroedinger}
\end{equation}
We thus want to find eigenvalues and normalizable eigenfunctions of
the rotated Hamiltonian
\begin{equation}
 H' = S_\theta H S^{-1}_\theta\,.
\end{equation}
The energies of the bound states are unaffected by the rotation,
whereas the continuum is rotated into the lower complex plane
by the angle $2\theta$ and the positions of the quasi-bound resonance
states are revealed.
For sufficiently large angles, these are independent of the value of $\theta$. 
Expressing $S_\theta \psi$ in a basis $\{\phi_i \}$, we obtain
\begin{equation}
 S_\theta|\psi \rangle = \sum_i c_i |\phi_i \rangle
\end{equation}
and
\begin{equation}
 \sum_i \langle \phi_j | S^{-1}_\theta H S^{-1}_\theta | \phi_i \rangle c_i = E \sum_i \langle \phi_j | S^{-1}_\theta S^{-1}_\theta | \phi_i \rangle c_i \,.
 \label{eq:GeneralizedEigenvalueProblemRotated}
\end{equation}
The solution can thus be obtained by using the rotated basis set
$\{S^{-1}_\theta|\phi_i\rangle \}$ with the unchanged Hamiltonian $H$.

%
We now express the wave function in an appropriate basis.
Our approach is identical to the one in Ref.~\cite{Zielinski2019},
i.e., for the radial part, we use the complete basis of Coulomb-Sturmian functions
\begin{equation}
  U_{NL}(\rho) = N_{NL}(2\rho)^L\mathrm{e}^{-\rho}L_N^{2L+1}(2\rho) \, ,
\label{eq:basis_r}
\end{equation}
with $\rho = r/\alpha$ and $N$ the radial quantum number.
Here, $\alpha$ is a free parameter.
For the angular part we use the spherical harmonics for $L$ with an
additional set of appropriate spin quantum numbers.
First, the quasispin $I$ and the hole spin $S_\mathrm{h}$ are coupled
to the effective hole spin $J$.
At the $\Gamma$ point, $J$ is a good quantum number and distinguishes
between the yellow ($J=1/2$) and green series ($J=3/2$).
Then $L$ and $J$ are coupled to the angular momentum $F$.
Finally, we take the electron spin into account by introducing
$F_{\mathrm{t}} = F + S_{\mathrm{e}}$ with the component
$M_{F_{\mathrm{t}}}$ along the quantization axis, which is chosen to
be the $[001]$ axis.
Our basis states thus are given by
\begin{equation}
  \ket{\Pi} = \ket{N,L;(I,S_\mathrm{h}),J;F,S_\mathrm{e};F_{\rm t},M_{F_{\rm t}}} \, .
\label{eq:basis}
\end{equation}
Using this basis set, we can transform the Schr\"odinger equation into the
generalized eigenvalue problem~\eqref{eq:GeneralizedEigenvalueProblemRotated},
which can be solved using the Lapack routine ZGGEV~\cite{lapackuserguide3}.
The complex-coordinate rotation is here achieved by rotating the free parameter
$\alpha$ $\rightarrow$ $\alpha =|\alpha|\mathrm{e}^{\mathrm{i}\theta}$.
Since the Coulomb-Sturmian functions are not orthogonal, the overlap
matrix in Eq.~\eqref{eq:GeneralizedEigenvalueProblemRotated} is
nontrivial~\cite{ImpactValence}.
To properly orthonormalize the eigenvectors, we apply a modified
Gram-Schmidt procedure.

%
Using the eigenfunctions obtained from the solution of the generalized
eigenvalue problem~\eqref{eq:GeneralizedEigenvalueProblemRotated}, we
are able to simulate absorption spectra for the yellow and green series.
The absorption coefficients are calculated with the formula~\cite{Zielinski2019}
\begin{equation}
  f(E) = -\frac{1}{\pi}{\mathrm{Im}}\sum_j\frac{f_{\rm rel}^{(j)}}{E-E_j}\,,
\label{eq:Spektrum}
\end{equation}
where $E_j$ are the complex energies of the resonance states and
\begin{equation}
  f_{\rm{rel}}\sim\left(\lim_{r \rightarrow 0}
    {\frac{\partial}{\partial r}\braket{\sigma^\pm_z|\Psi(\boldsymbol{r})}}\right)^2 
\label{eq:frel}
\end{equation}
is the complex generalization of the relative oscillator strength.
The overlaps with the states
\begin{align}
|\sigma^+_z\rangle &= |2,-1\rangle_D, \nonumber\\
|\sigma^-_z\rangle &= -|2,1\rangle_D,
\end{align}
determine the spectrum for $\sigma^+$ and $\sigma^-$ polarized light.
Here, we use the abbreviation
\begin{align}
  \ket{F_{\rm t},M_{F_{\rm t}}}_D
  &= \ket{(S_{\rm e},S_{\rm h})\,S,\,I;\,I+S,\,L;\,F_{\rm t},\,M_{F_{\rm t}}}\nonumber\\
  &=\ket{(1/2,1/2)\,0,\,1;\,1,\,1;\,F_{\rm t},\,M_{F_{\rm t}}} \, ,
\end{align}
to denote the states with a coupling scheme differing from the one in
the basis states.

\section{Results and discussion}
\label{sec:results}
The results for both the yellow and green exciton series are presented
in Fig.~\ref{fig:SpectrumResonances}.
In the computations we have used the basis set \eqref{eq:basis} with
$N + L < 50$, $|\alpha| = 63$, and $\theta = 0.14$ and restricted
ourselves to the odd states, as only those contribute to the
absorption coefficient of one-photon transitions.
In the lower part of Fig.~\ref{fig:SpectrumResonances} we show the
resonance positions of the yellow and green exciton series in the
complex energy plane.
Clearly visible are the bound states of the yellow exciton series at
energies below the gap energy $E_{\mathrm{g}}$ and the resonances of
the green exciton series at energies below the band edge
$E_{\mathrm{g}}+\Delta$.
Above the band edges energies are bundled along straight lines and
rotated into the complex plane.
The rotation angle is nearly given by $2\theta$ as is expected for
complex rotated continuum states~\cite{Reinhardt1982,Ho83,MOISEYEV1998}.
Note that the numerical resonance positions are already rotated into
the lower complex energy plane at energies slightly below the band
edges.
This is a numerical artifact due to the finite size of the basis set.

The upper part of Fig.~\ref{fig:SpectrumResonances} presents the
corresponding absorption spectrum obtained with Eq.~\eqref{eq:Spektrum}.
Since we do not include the effects of phonons in our model, the
yellow exciton states here are bound states with infinite lifetimes.
To avoid $\delta$ function type absorption peaks we have simulated
the interaction with phonons by manually introducing the finite
linewidths
\begin{equation}
  \gamma_n = 9\,\mathrm{meV}(n^2-1)/n^5
  \label{eq:yellowLinewidths}
\end{equation}
with an effective principal quantum number
\begin{equation}
  n = \sqrt{E_{\mathrm{Ryd}}/(E_{\mathrm{g}} - E)} + \delta_P
\end{equation}
derived from the approximate Rydberg formula~\cite{GiantRydbergExcitons,Stolz2018}
with $E_\mathrm{Ryd} = 92\,\mathrm{meV}$ and $\delta_P = 0.23$.

The green exciton states, however, are true resonances even apart from
phonons, since they are coupled to the yellow continuum by the valence
band structure.
The linewidths visible in the upper part of Fig.~\ref{fig:SpectrumResonances}
are solely due to this effect.
The continuum states of both the yellow and green excitons provide for
a square root function shaped background starting at energies above
the respective band edges.

We now want to discuss the classification and symmetries of resonances
of the green exciton series.
The green excitons are defined by the condition $J = 3/2$.
Additionally, we have to consider the angular momentum $L$.
The electron spin plays no role for the odd states and remains a good
quantum number.
Thus, for the P states we have $F = 1/2$, $3/2$, and $5/2$.
Reducing the symmetry to the octahedral group $O_{\mathrm{h}}$,
the irreducible representations are $\Gamma_6^-$, $\Gamma_8^-$, and
$\Gamma_7^- \oplus \Gamma_8^-$, respectively~\cite{koster1963properties}.
We have to take into account, however, that the quasispin
$\boldsymbol{I}$ transforms according to $\Gamma_5^+$ instead of
$\Gamma_4^+$.
Since $\Gamma_5^+ = \Gamma_4^+ \otimes \Gamma_2^+$, this can be done by
performing the coupling of angular momenta as usual, but multiplying
by $\Gamma_2^+$ in the end.
For the P states, we thus have the representations  $\Gamma_6^-$,
$\Gamma_8^-$, $\Gamma_7^-$, and $\Gamma_8^-$.
Note that the degeneracy of these states is doubled due to the
electron spin.
Since half-integer angular momenta only have the above mentioned irreducible
representations in $O_\mathrm{h}$~\cite{koster1963properties}, all states considered in this work
can be classified according {to} them.
Of those, only $\Gamma_6^-$ and $\Gamma_8^-$ states are dipole active,
because only those contain $\Gamma_4^-$ when multiplied with the electron spin
symmetry $\Gamma_6^+$~\cite{koster1963properties}.
Additional consideration of the degree of degeneracy then allows for
the unique assignment of irreducible representations to the odd
exciton states as given in the Supplemental
Material~\cite{RommelGruen2019Supplemental}.

\begin{table}[b]
  \caption{Numerically determined resonance positions of some of the
    lowest P states belonging to the irreducible representation
    $\Gamma_6^-$.  The selected states produce the dominant peak of
    each $n$-manifold in the absorption spectrum.}
  \begin{tabularx}{\columnwidth}
    {r@{\extracolsep{\fill}}c@{\extracolsep{\fill}}c@{\extracolsep{\fill}}c@{\extracolsep{\fill}}c@{\extracolsep{\fill}}}
    \toprule
    State & $\mathrm{Re}\,E~[\mathrm{eV}]$ & $\mathrm{Im}\,E~[\mathrm{meV}]$&	Re $f_{\mathrm{rel}}$ &	Im $f_{\mathrm{rel}}$\\ 
\hline
 2P  & 2.26887 &	-3.01965 &      \phantom{-}4.2998	& \phantom{-}5.8604 \\
 3P  & 2.28765 &	-0.90691 &	\phantom{-}1.1603	& \phantom{-}1.8982 \\
 4P  & 2.29423 &	-0.38575 &	\phantom{-}0.5028	& \phantom{-}0.8270 \\
 5P  & 2.29731 &	-0.19095 &	\phantom{-}0.2571	& \phantom{-}0.4496 \\
 6P  & 2.29901 &	-0.10700 &	\phantom{-}0.1337	& \phantom{-}0.2630 \\
 7P  & 2.30005 &	-0.06821 &	\phantom{-}0.1416	& \phantom{-}0.1579 \\
 8P  & 2.30072 &	-0.04328 &	\phantom{-}0.0519	& \phantom{-}0.1085 \\
 9P  & 2.30120 &	-0.03314 &	\phantom{-}0.0438	& \phantom{-}0.0730 \\
 10P & 2.30154 &	-0.02334 &	\phantom{-}0.0302	& \phantom{-}0.0550 \\
 11P & 2.30180 &	-0.01733 &	\phantom{-}0.0268	& \phantom{-}0.0387 \\
 12P & 2.30199 &	-0.01459 &	\phantom{-}0.0198	& \phantom{-}0.0322 \\
 13P & 2.30215 &	-0.01012 &	\phantom{-}0.0168	& \phantom{-}0.0223 \\
 14P & 2.30227 &	-0.00903 &	\phantom{-}0.0118	& \phantom{-}0.0212 \\
 15P & 2.30237 &	-0.00742 &	\phantom{-}0.0105	& \phantom{-}0.0165 \\
      \hline
     \end{tabularx}
\label{tab:ResonancePositions}
\end{table}

\begin{figure}
 \includegraphics[width=1.0\columnwidth]{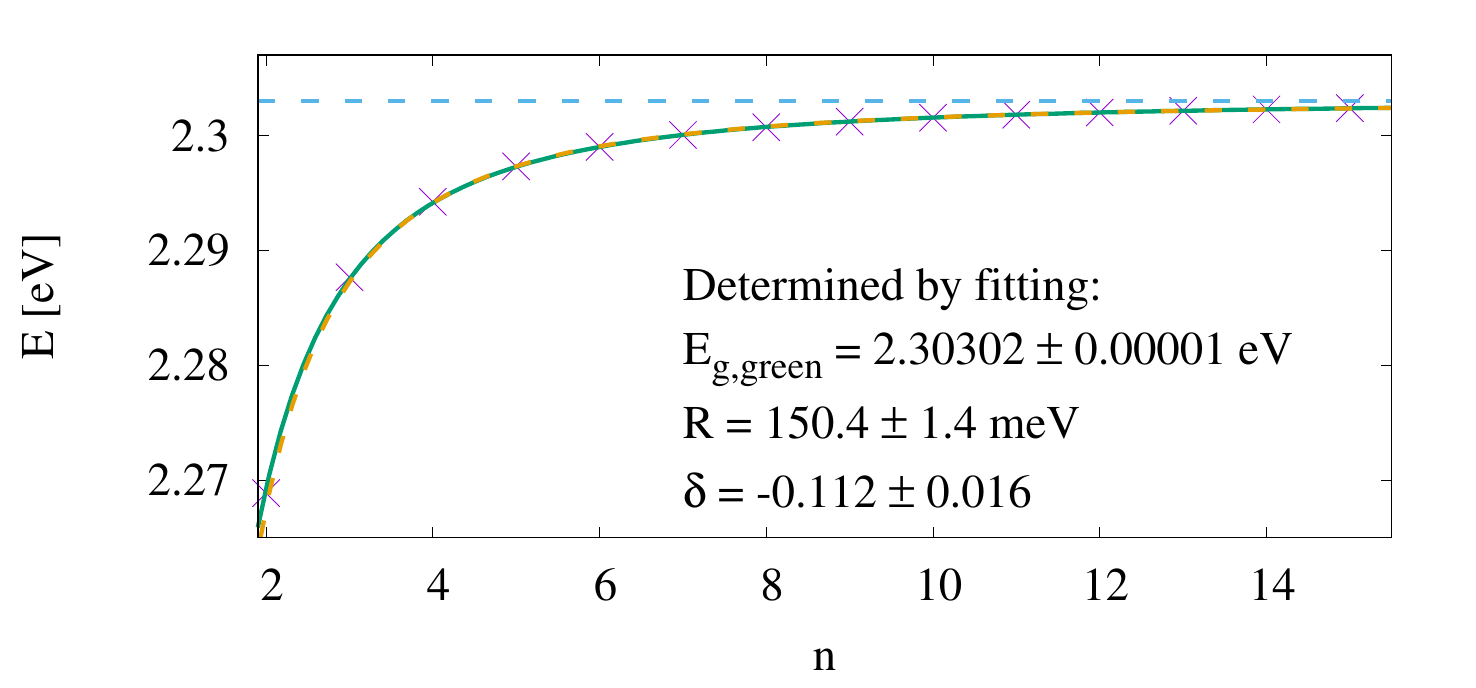}
 \caption{Numerically determined energies of the lowest dominant P states.  The 
   band gap energy $E_{\mathrm{g,green}}$, Rydberg constant $R$, and
   quantum defect $\delta$ are determined by a fit {(green solid line). For comparison, the dashed yellow
   line shows a fit without quantum defect (see text).}
   \label{fig:EgapRydDel}}
\end{figure}

In Table~\ref{tab:ResonancePositions} and in Fig.~\ref{fig:EgapRydDel}
the resonance positions of the dominant P states are presented. We extract the band gap, Rydberg constant and quantum defect using a
fit of the form $E(n) = E_\mathrm{g} - R/(n - \delta)^2$.
As expected, the fitted continuum threshold
$E^{\mathrm{fit}}_{\mathrm{g,green}}=2.30302\,\mathrm{eV}$ shows
excellent agreement with the band gap of the green excitons
$E_{\mathrm{g}} + \Delta = 2.30308\,\mathrm{eV}$.
For the Rydberg constant we obtain $R = 150.4\,\mathrm{meV}$, which is
in good agreement with
literature~\cite{Schoene2016,Uihlein1981,Malerba2011,Grun1961}.
{In previous theoretical work by Sch\"one \emph{et al.}~\cite{Schoene2016}, the quantum defect of the green exciton series was investigated using a simplified treatment of the valence band dispersion neglecting the coupling of the
green resonances to the yellow continuum, yielding negative quantum defects, which for the P states are in reasonable agreement with our result of $\delta = -0.112$.}
{In Fig.~\ref{fig:EgapRydDel} we also present a fit without using a quantum defect. Detailed comparison shows that this fit is slightly less accurate,
especially for low principal quantum numbers. This motivates the validity of the quantum-defect corrected
Rydberg formula also for the dominant P states of the green exciton series.}
A more complete version of Table~\ref{tab:ResonancePositions}
is given in the  Supplemental Material~\cite{RommelGruen2019Supplemental}.

{The green exciton series has already been experimentally investigated in Ref.~\cite{Gross1956}. At temperature $T = 4.2\,\mathrm{K}$ Gross found
$E_{2\mathrm{P}} = 2.266\,\mathrm{eV}$, $E_{3\mathrm{P}} = 2.287\,\mathrm{eV}$, $E_{4\mathrm{P}} = 2.294\,\mathrm{eV}$
and $E_{5\mathrm{P}} = 2.298\,\mathrm{eV}$, which agrees with the numerical energies given in Table~\ref{tab:ResonancePositions} to within approximately $1$ - $2\,\mathrm{meV}$.}

\begin{figure}
 \includegraphics[width=1.0\columnwidth]{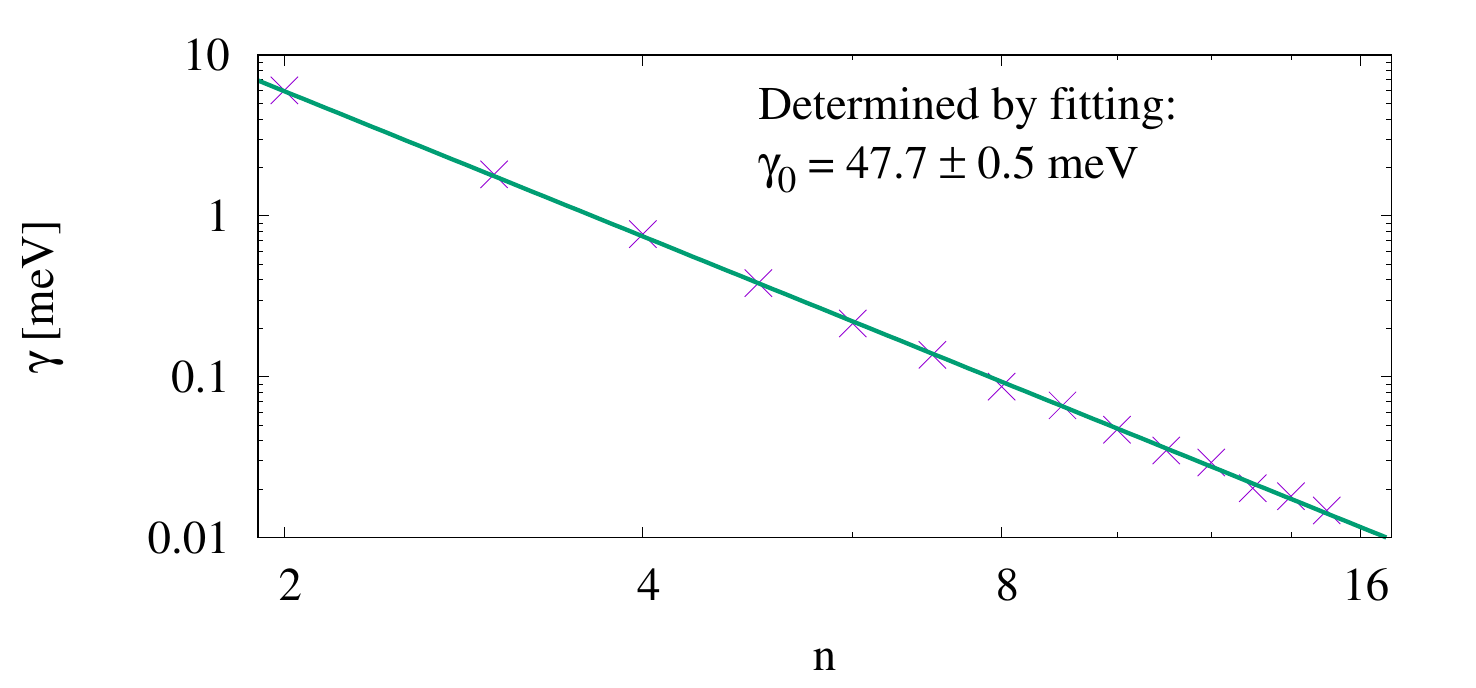}
 \caption{The {numerically determined} yellow-continuum induced linewidths of the dominant
   green P states. A fit using the function $\gamma_n = \gamma_0 n^{-3}$
   is in good agreement with the {numerical} data.
   \label{fig:FitLinewidths}}
\end{figure}
The behavior of the linewidths of the green excitons induced by the
coupling to the yellow continuum as a function of the principal
quantum number is shown in Fig.~\ref{fig:FitLinewidths}.
A function of the form $\gamma_n = \gamma_0 n^{-3}$ provides a good fit to
the numerically determined values. We obtain $\gamma_0 = 47.7\,\mathrm{meV}${,
which means that the
yellow-continuum induced linewidths of the dominant green P states
are large compared to the phonon-coupling induced linewidths of the
yellow excitons given in Eq.~\eqref{eq:yellowLinewidths}.  Assuming that the phonon-coupling
leads to similar linewidths for both the yellow and green excitons,
the widths of the green excitons shown in Fig.~\ref{fig:SpectrumResonances} would only
slightly increase when taking phonon-coupling into account, however,
a more detailed theory of the phonons or precise state-of-the-art
experimental data are necessary to clarify this point.}

\section{Conclusion and outlook}
\label{sec:conclusion}
We have computed the resonance positions, linewidths, and relative
oscillator strengths of the green exciton series of cuprous oxide,
thereby taking into account the valence band structure of the crystal
and the coupling of the green excitons to the yellow continuum.
For the computations we have used a complete basis set with
Coulomb-Sturmian functions for the radial part of the wave function
and the complex-coordinate rotation method.
For the dominant P states in the absorption spectrum we have confirmed
their hydrogen-like behavior and extracted the Rydberg energy and
quantum defect, which are in good agreement with
literature~\cite{Schoene2016}.
The linewidths of the green P states decrease $\sim n^{-3}$ with
increasing principal quantum number.

In Sec.~\ref{sec:results} we have compared some resonance positions to
the experimental work of Gross~\cite{Gross1956}. In the meantime
experimental techniques have made substantial progress. A comparison
with new data would thus be desirable.
The interesting question is whether giant Rydberg states of the green
exciton series with quantum numbers up to $n\approx 25$ and the
computed fine structure splitting can be experimentally observed
similar as for the yellow
series~\cite{GiantRydbergExcitons,ObservationHighAngularMomentumExcitons}.

In this paper, we have focused on the odd states.
The 1S state of the even green series is bound and has been computed,
including the central-cell corrections, in
Ref.~\cite{frankevenexcitonseries}.
In the future we can also investigate the even resonance states of the
green exciton series.

Interseries transitions are currently investigated~\cite{KruegerInterseries2019}.
Starting from the present investigations, we can now go on to calculate the
interseries transition amplitudes between the yellow and green series,
taking the valence band structure into account.

\acknowledgments
This work was supported by Deutsche Forschungsgemeinschaft (DFG)
through Grant No.~MA1639/13-1.
We thank Frank Schweiner for his contributions and G\"unter Wunner for
a careful reading of the manuscript.


%


%
%
%
%
%
%

\section*{Supplemental material} 
%
%

\setcounter{equation}{0}
\setcounter{figure}{0}
\setcounter{table}{0}
In this Supplemental Material we will present additional numerical
data for the green exciton series, including an assignment of
approximate quantum numbers, group theoretical representations and
degeneracies of states.
The assignment of approximate quantum numbers is a nontrivial task,
due to the strong overlap of $n$ and $L$ manifolds.
To this end, starting with the hydrogen-like exciton model, we slowly
switch on the band structure and follow the resonance positions.
We also apply projection operators to the eigenstates $|\psi \rangle$
to help with the assignment of quantum numbers.

The Hamiltonian of excitons in cuprous oxide in relative coordinates
can be separated into the
form (see Refs.~\cite{Luttinger52CyclotronResonanceSemiconductors,ImpactValence} in the paper).
\begin{equation}
  H = E_\mathrm{g} + H_\mathrm{0}+ H_\mathrm{SO} + \lambda H_\mathrm{vb}\,,
  \label{eq:SeparatedHamiltonian}
\end{equation}
with the hydrogen-like part
\begin{equation}
  H_\mathrm{0} = \frac{\gamma_1'\boldsymbol{p}^2}{2m_0} -\frac{e^2}{4\pi\varepsilon_0\varepsilon|\boldsymbol{r}|}\,,
\end{equation}
using $\gamma_1' = \gamma_1 + m_0/m_\mathrm{e}$, the spin-orbit coupling term
\begin{equation}
 H_{\rm SO}=\frac{2}{3}\Delta
 \left(1+\frac{1}{\hbar^2}\boldsymbol{I}\cdot\boldsymbol{S}_{\rm h}\right)\,,
\end{equation}
and additional terms stemming from the complex valence band structure $H_\mathrm{vb}$,
%
\begin{align}
  H_\mathrm{vb} &= \frac{1}{2\hbar^2m_0}
  \{4\hbar^2\gamma_2\boldsymbol{p}^2
 +2(\eta_1+2\eta_2)\boldsymbol{p}^2(\boldsymbol{I}\cdot\boldsymbol{S}_{\rm h})\nonumber\phantom{\frac{1}{2}}\\
&-6\gamma_2(p^2_1\boldsymbol{I}^2_1+{\rm c.p.})
-12\eta_2(p^2_1\boldsymbol{I}_1\boldsymbol{S}_{\rm h1}+{\rm c.p.})\nonumber\phantom{\frac{1}{2}}\\
&-12\gamma_3(\{p_1,p_2\}\{\boldsymbol{I}_1,\boldsymbol{I}_2\}+{\rm c.p.})\nonumber\phantom{\frac{1}{2}}\\
\phantom{\frac{1}{2}} &-12\eta_3(\{p_1,p_2\}(\boldsymbol{I}_1\boldsymbol{S}_{\rm h2}
         +\boldsymbol{I}_2\boldsymbol{S}_{\rm h1})+{\rm c.p.})\}\,.
\end{align}

\begin{figure*}[b]
  \includegraphics[width=0.9\textwidth]{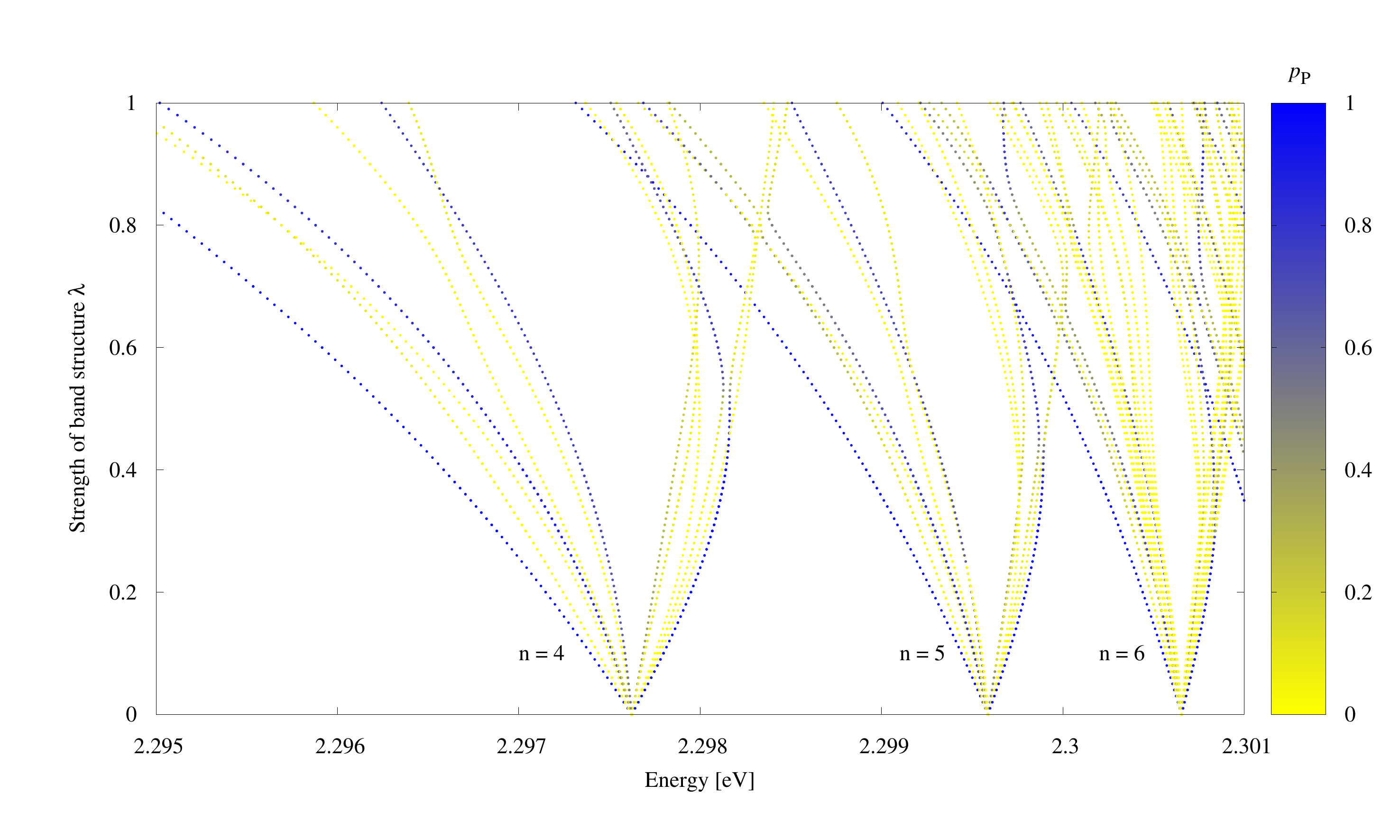}
  \includegraphics[width=0.9\textwidth]{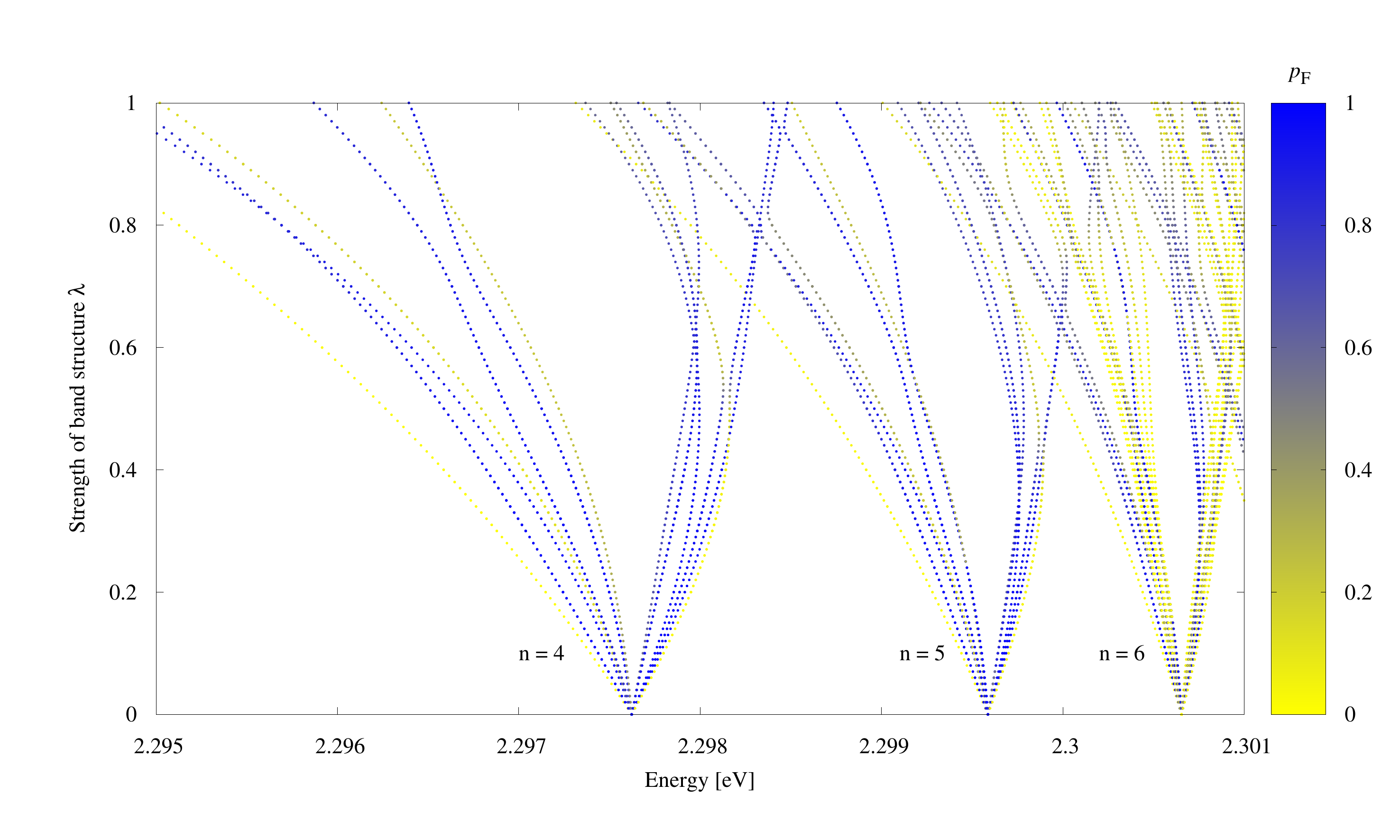}
  \caption{Energies of states as a function of the strength of the band
    structure. A value of $\lambda = 0$ means that the band structure is
    completely switched off, whereas a value of $\lambda = 1$
    signifies that the band structure is completely switched on. The
    color palette shows the P state component $p_\mathrm{P}$ (top) and
    F state component $p_\mathrm{F}$ (bottom) given in
    Eqs.~\eqref{eq:OverlapP} and \eqref{eq:OverlapF}, respectively.}
 \label{fig:Bandstructure}
 \end{figure*}

To make possible the assignment of states, we introduce the parameter
$\lambda$ in Eq.~\eqref{eq:SeparatedHamiltonian} controlling the
strength of the band structure. For $\lambda = 0$, the excitons are
described in the hydrogen-like model and with $\lambda = 1$ the full
band structure is switched on.
We now diagonalize the Hamiltonian as described in the Paper, while
varying $\lambda$ between zero and one.
Following the resonance states in the resulting $E$-$\lambda$-diagram
then allows for the assignment of the principal quantum number as
shown in Fig.~\ref{fig:Bandstructure}. Still, the assignment remains
ambiguous due to the existence of various avoided crossings.
We additionally calculate the P state component
\begin{equation}
  p_\mathrm{P} = |\langle L = 1 | \psi \rangle|^2 = \langle \psi | P_{L=1} | \psi \rangle
  \label{eq:OverlapP}
\end{equation}
and the F state component
\begin{equation}
  p_\mathrm{F} = |\langle L = 3 | \psi \rangle|^2 = \langle \psi | P_{L=3} | \psi \rangle
  \label{eq:OverlapF}
\end{equation}
with the corresponding projection operators $P_{L=1}$ and $P_{L=3}$, respectively.

In Table~\ref{tab:ResonancePositions} we present the numerical data
for all odd parity green resonance states with energies up to
$E = 2.299\,\mathrm{eV}$ and dominant P states up to $n=15$.
The table includes the assignment of the approximate principal quantum
number $n$, orbital angular momentum $L$, irreducible representation,
degeneracy $g$, complex resonance energy $E$, and the complex relative
oscillator strength $f_{\mathrm{rel}}$.
Furthermore, we provide the values for $p_\mathrm{P}$ and
$p_\mathrm{F}$ computed according to Eqs.~\eqref{eq:OverlapP} and
\eqref{eq:OverlapF} and the yellow admixture
\begin{equation}
  p_\mathrm{y} = |\langle J = 1/2 | \psi \rangle|^2 = \langle \psi | P_{J=1/2} | \psi \rangle\,.
  \label{eq:OverlapY}
\end{equation}

\begin{table*}[b]
      \caption{Numerical data for all odd parity green resonance
        states with energies up to $E = 2.299\,\mathrm{eV}$ and
        dominant P states up to $n=15$. The table includes the
        assignment of the approximate principal quantum number $n$,
        orbital angular momentum $L$, irreducible representation,
        degeneracy $g$, complex resonance energy $E$, and the complex
        relative oscillator strength $f_{\mathrm{rel}}$. Furthermore,
        we provide the values for $p_\mathrm{P}$ and $p_\mathrm{F}$
        computed according to Eqs.~\eqref{eq:OverlapP} and
        \eqref{eq:OverlapF} and the yellow admixture $p_\mathrm{y}$
        according to Eq.~\eqref{eq:OverlapY}. For one of the $4$F
        states, we could not get a precise value for $p_\mathrm{P}$
        due to convergence problems.}
      \begin{tabularx}{\textwidth}
    {r@{\extracolsep{\fill}}c@{\extracolsep{\fill}}c@{\extracolsep{\fill}}c@{\extracolsep{\fill}}c@{\extracolsep{\fill}}c@{\extracolsep{\fill}}cccc}
    \toprule
State            &  Irrep.  & $g$ & $\mathrm{Re}\,E~[\mathrm{eV}]$ & $\mathrm{Im}\,E~[\mathrm{meV}]$ & 	Re $f_\mathrm{rel}$ &	Im $f_\mathrm{rel}$ &  $p_{\mathrm P}$ & $p_{\mathrm F}$ & $p_{\mathrm y}$ \\
\hline
{2P}      &		$\Gamma_6^-$ &4 &	   2.26887 &	-3.01965 &  \phantom{-}4.2998	& \phantom{-}5.8604	&       0.955 &	0.043    &	0.018 \\
{2P}      &		$\Gamma_8^-$ &8 &	   2.27404 &	-1.19593 &  \phantom{-}0.7471	& \phantom{-}1.4121 	&       0.879 &	0.117    &	0.010 \\
{2P}      &		$\Gamma_7^-$ &4 &	   2.27948 &	-0.81870 &  \phantom{-}0.0000	& \phantom{-}0.0000     &       0.856 &	0.136    &	0.017 \\
{2P}      &		$\Gamma_8^-$ &8 &	   2.28559 &	-0.04207 &            -0.0177   & \phantom{-}0.0146     &       0.857 &	0.141    &	0.004 \\
{3P}      &		$\Gamma_6^-$ &4 &	   2.28765 &	-0.90691 &  \phantom{-}1.1603	& \phantom{-}1.8982     &	0.966 &	0.030    &	0.009 \\
{3P}      &		$\Gamma_8^-$ &8 &	   2.28944 &	-0.39764 &  \phantom{-}0.2350	& \phantom{-}0.4813     &	0.895 &	0.099    &	0.004 \\
{3P}      &		$\Gamma_7^-$ &4 &	   2.29178 &	-0.26468 &  \phantom{-}0.0000	& \phantom{-}0.0000     &	0.875 &	0.118    &	0.007 \\
{3P}      &		$\Gamma_8^-$ &8 &	   2.29384 &	-0.00734 &            -0.0052	& \phantom{-}0.0083     &	0.611 &	0.379    &	0.001 \\
{4P}      &		$\Gamma_6^-$ &4 &	   2.29423 &	-0.38575 &  \phantom{-}0.5028	& \phantom{-}0.8270     &	0.941 &	0.050    &	0.004 \\
{4F}      &		$\Gamma_6^-$ &4 &	   2.29474 &	-0.00149 &  \phantom{-}0.0039	&           -0.0007     &	0.030 &	0.833    &	0.001 \\
{4F}      &		$\Gamma_8^-$ &8 &	   2.29486 &	-0.04633 &            -0.0475	& \phantom{-}0.0259     &	0.030 &	0.842    &	0.002 \\
{4P}      &		$\Gamma_8^-$ &8 &	   2.29502 &	-0.13785 &  \phantom{-}0.1574	& \phantom{-}0.1803     &	0.885 &	0.110    &	0.001 \\
{4F}      &		$\Gamma_8^-$ &8 &	   2.29587 &	-0.00089 &            -0.0001	& \phantom{-}0.0001     &	0.012 &	0.824    &	0.001 \\
{4P}      &		$\Gamma_7^-$ &4 &	   2.29625 &	-0.12115 &  \phantom{-}0.0000	& \phantom{-}0.0000     &	0.769 &	0.222    &	0.003 \\
{4F}      &		$\Gamma_8^-$ &8 &	   2.29639 &	-0.00593 &            -0.0024	& \phantom{-}0.0131     &	0.072 &	0.905    &	0.001 \\
{5P}      &		$\Gamma_6^-$ &4 &	   2.29731 &	-0.19095 &  \phantom{-}0.2571	& \phantom{-}0.4496     &	0.905 &	0.091    &	0.002 \\
{4F}      &		$\Gamma_6^-$ &4 &	   2.29737 &	-0.00676 &  \phantom{-}0.0004	&           -0.0169     &    $\sim 0$ & 0.565    &      0.000 \\
{4P}      &		$\Gamma_8^-$ &8 &	   2.29751 &	-0.00784 &            -0.0061	& \phantom{-}0.0080     &	0.461 &	0.476    &	0.001 \\
{4F}      &		$\Gamma_8^-$ &8 &	   2.29754 &	-0.00193 &            -0.0003	& \phantom{-}0.0025     &	0.058 &	0.550    &	0.000 \\
{5F}      &		$\Gamma_6^-$ &4 &	   2.29766 &	-0.00303 &  \phantom{-}0.0092	& \phantom{-}0.0002     &	0.095 &	0.788    &	0.001 \\
{5P}      &		$\Gamma_8^-$ &8 &	   2.29769 &	-0.08012 &  \phantom{-}0.0503	& \phantom{-}0.0934     &	0.748 &	0.239    &	0.001 \\
{4F}      &		$\Gamma_7^-$ &4 &	   2.29782 &	-0.00014 &  \phantom{-}0.0000	& \phantom{-}0.0000     &	0.035 &	0.622    &	0.000 \\
{5F}      &		$\Gamma_8^-$ &8 &	   2.29783 &	-0.01073 &  \phantom{-}0.0158	& \phantom{-}0.0061     &	0.313 &	0.596    &	0.000 \\
{5F}      &		$\Gamma_8^-$ &8 &	   2.29835 &	-0.00059 &            -0.0001	& \phantom{-}0.0000     &	0.013 &	0.835    &	0.001 \\
{4F}      &		$\Gamma_7^-$ &4 &	   2.29841 &	-0.01880 &  \phantom{-}0.0000	& \phantom{-}0.0000     &	0.035 &	0.896    &	0.001 \\
{4F}      &		$\Gamma_8^-$ &8 &	   2.29848 &	-0.00070 &  \phantom{-}0.0006	& \phantom{-}0.0009     &	0.097 &	0.846    &	0.000 \\
{5P}      &		$\Gamma_7^-$ &4 &	   2.29851 &	-0.04862 &  \phantom{-}0.0000	& \phantom{-}0.0000     &	0.791 &	0.175    &	0.001 \\
{5F}      &		$\Gamma_8^-$ &8 &	   2.29875 &	-0.00545 &            -0.0017	& \phantom{-}0.0127     &	0.090 &	0.897    &	0.001 \\
{6P}      &		$\Gamma_6^-$ &4 &	   2.29901 &	-0.10700 &  \phantom{-}0.1337	& \phantom{-}0.2630     &	0.812 &	0.164    &	0.001 \\
{7P}      &	  $\Gamma_6^-$ &4 &    2.30005 &	-0.06821 &	\phantom{-}0.1416	& \phantom{-}0.1579     &	0.671 &	0.275    &	0.001 \\
{8P}      &	  $\Gamma_6^-$ &4 &    2.30072 &	-0.04328 &	\phantom{-}0.0519	& \phantom{-}0.1085     &	0.648 &	0.290    &	0.001 \\
{9P}      &	  $\Gamma_6^-$ &4 &    2.30120 &	-0.03314 &	\phantom{-}0.0438	& \phantom{-}0.0730     &	0.510 &	0.368    &	0.000 \\
{10P}     &	  $\Gamma_6^-$ &4 &    2.30154 &	-0.02334 &	\phantom{-}0.0302	& \phantom{-}0.0550     &	0.509 &	0.383    &	0.000 \\
{11P}     &	  $\Gamma_6^-$ &4 &    2.30180 &	-0.01733 &	\phantom{-}0.0268	& \phantom{-}0.0387     &	0.455 &	0.429    &	0.000 \\
{12P}     &	  $\Gamma_6^-$ &4 &    2.30199 &	-0.01459 &	\phantom{-}0.0198	& \phantom{-}0.0322     &	0.363 &	0.418    &	0.000 \\
{13P}     &	  $\Gamma_6^-$ &4 &    2.30215 &	-0.01012 &	\phantom{-}0.0168	& \phantom{-}0.0223     &	0.404 &	0.462    &	0.000 \\
{14P}     &	  $\Gamma_6^-$ &4 &    2.30227 &	-0.00903 &	\phantom{-}0.0118	& \phantom{-}0.0212     &	0.327 &	0.408    &	0.000 \\
{15P}     &	  $\Gamma_6^-$ &4 &    2.30237 &	-0.00742 &	\phantom{-}0.0105	& \phantom{-}0.0165     &	0.321 &	0.474    &	0.000 \\
\hline
     \end{tabularx}
  \label{tab:ResonancePositions}
\end{table*}

\end{document}